\input harvmac
\overfullrule=0pt

%
\def\t{{\theta}}
\def\s{{\sigma}}
\def\e{{\epsilon}} 
\def\hhalf{{\scriptstyle{1\over 2}}}
\def\N{{\nabla}}

\def\p{{\partial}}

\def\bar{\overline}
\def\R{{\bf R}}
\def\T{{\bf T}}
\def\K3{{\bf K3}}
\def\AdS{{\bf AdS}}

\Title{ \vbox{\baselineskip12pt
\hbox{hep-th/9910205}
\hbox{IASSNS-HEP-99/66}
\hbox{}}}
{\vbox{\centerline{Vertex Operators for AdS3 Background}
\bigskip
\centerline{  With Ramond-Ramond Flux }}}
\smallskip
\centerline{Louise Dolan}
\smallskip
\centerline{\it Department of Physics}
\centerline{\it
University of North Carolina, Chapel Hill, NC 27599-3255}
\bigskip
\smallskip
\centerline{Edward Witten}
\smallskip
\centerline{\it California Institute of Technology, Pasadena, CA 91125, USA}
\centerline{\it CITUSC Center For Theoretical Physics}
\centerline{\it and}
\centerline{\it School of Natural Sciences, Institute for Advanced Study}
\centerline{\it Olden Lane, Princeton, NJ 08540, USA}\bigskip
\def\sqr#1#2{{\vbox{\hrule height.#2pt\hbox{\vrule width
.#2pt height#1pt \kern#1pt\vrule width.#2pt}\hrule height.#2pt}}}
\def\Box{\mathchoice\sqr64\sqr64\sqr{4.2}3\sqr33}

\bigskip

\noindent
In order to study vertex operators for the Type IIB superstring on 
${\bf AdS}$ space,
we derive supersymmetric constraint equations for the vertex operators
in ${\bf AdS}_3\times {\bf S}^3$ backgrounds with Ramond-Ramond
flux, using Berkovits-Vafa-Witten variables. These constraints are solved
to compute the vertex operators and show that they
satisfy the linearized $D=6$, $N=(2,0)$ equations
of motion for a supergravity and tensor multiplet expanded around the 
${\bf AdS}_3\times {\bf S}^3$ spacetime.  
 
\Date{}

\lref\bvw{N. Berkovits, C. Vafa, and E. Witten, ``Conformal Field Theory
of AdS Background With Ramond-Ramond Flux'', JHEP 9903: 018, 1999;
hep-th/9902098.}

\lref\bv{N. Berkovits and C. Vafa, ``$N=4$ Topological Strings'',
Nucl. Phys. B433 (1995) 123, hep-th/9407190.}

\lref\b{N. Berkovits, ``A New Description of the Superstring'',
Jorge Swieca Summer School 1995, p. 490, hep-th/9604123.}

\lref\r{L. Romans, ``Self-Duality for Interacting Fields'',
Nucl. Phys. B276 (1986) 71.}

\lref\gsw{M. Green, J. Schwarz, and E. Witten, 
{\it Superstring~  Theory}, vol. 2, p. 269, Cambridge University Press: 
Cambridge, U.K. 1987.} 

\lref\sez{S. Deger, A. Kaya, E. Sezgin and P. Sundell,
``Spectrum of $D=6, N=4b$ Supergravity on $AdS_3 \times S^3$'',
Nucl. Phys. {\bf B536} (1988) 110; hepth/9804166.}

\lref\btwo{N. Berkovits, ``Quantization of the Type II Superstring
in a Curved Six-Dimensional Background'', hep-th/9908041.}


\def\S{{\bf S}}
\def\T{{\bf T}}
\def\AdS{{\bf AdS}}
\def\T{{\bf T}}

\newsec{Introduction}

The formulation of compactified string theories on
anti de Sitter (AdS) backgrounds is necessary to understand
the conjectured dualities with spacetime conformal field theories (CFT's).
Presently, calculations using 
the correspondence are restricted
to the supergravity limit of the string theories. 
This is because of the appearance of
background Ramond-Ramond fields in the string worldsheet action, which
 makes understanding the worldsheet CFT's difficult.

This problem has been partly overcome in some special cases.  For example,
the Berkovits-Vafa formalism for supersymmetric quantization on $\R^6\times
M$ (where $M$ is $\K3$ or $\T^4$)
\refs{\bv,\b,\btwo}  uses the following worldsheet fields.
In addition to bosonic fields $x^p(z,\bar z)$ that contain
both left- and right-moving modes, there are left-moving fermi fields
$ \t^a_L(z),
p^a_L(z)$ of spins 0 and 1, together with ghosts
$\s_L(z),\rho_L(z)$, and right-moving counterparts of all
these left-moving fields.  The virtue of this formalism
is that Ramond-Ramond background fields can be incorporated
without adding spin fields to the worldsheet action.   In the $\AdS_3
\times \S^3$ case, after adding those backgrounds it becomes
convenient \bvw\ to integrate out the $p$'s, giving a model
in which the fields are $x^p$, $\t^a$, $\bar\t^a$ (all now with
both left- and right-moving components) as well as the ghosts.
$x^p$, $\t^a$, and $\bar\t^a$ can be combined together as
coordinates on a supergroup manifold $PSU(2|2)$.\foot{$PSU(2|2)$ is
the projectivization of $SU(2|2)$, obtained by dividing by its center,
which is $U(1)$.  $PSU(2|2)$ has been called $SU'(2|2)$ in \bvw.}
The model is thus
a sigma model with that supergroup as a target, coupled also
to the ghosts $\rho$ and $\sigma$.  The spacetime supersymmetry
group is $PSU(2|2)\times PSU(2|2)$, acting by left and right 
multiplication on the $PSU(2|2)$ manifold, that is, by $g\to agb^{-1}$
where $g$ is a $PSU(2|2)$-valued field (which combines $x,\theta, $
and $\bar\theta$), and $a,b\in PSU(2|2)$ are the symmetry group
elements.

Thus one arrives \bvw\ at a sigma model with conventional local
interactions (no spin fields in the Lagrangian) that gives
a conformal field theory description of strings propagating
on $\AdS_3\times \S^3$, with  manifest spacetime supersymmetry.
This gives a framework for studying
 the type IIB superstring on $\AdS_3\times \S^3\times M$,
 and the model can in principle be used
to go beyond the supergravity approximation
and compute the $\alpha'$ expansion of correlation functions. 

In this paper, we analyze the vertex operators for this model
for the massless states that are independent of the details
of the compactification (including the supergravity multiplet).
We work to leading order in $\alpha'$, but because of the high
degree of symmetry of the model, it seems very likely that the result is
exact.  Concretely, we look for vertex operators that
are constructed from a function of the group-valued fields $g$
together with the ghosts $\rho,\sigma$ (but with no dependence
on derivatives of the fields).
 The motivation for this assumption is that it holds
in the flat case, according to \bv.

The discussion is organized as follows.
In sect. 2, we write the string constraints and
find an explicit description of
the action of the $PSU(2|2)^2$ generators
on the vertex operators.
We construct the constraint equations for the vertex operators
in terms of these charges, by requiring the equations to be
a smooth deformation of the known flat space case and to
 be invariant under the AdS supersymmetry transformations.
 In sect. 3, we  discuss how the gauge transformations 
of the vertex operator equations distinguish the physical 
degrees of freedom. In sect. 4 we evaluate the constraints,
using the differential geometry of the group manifold. 
In sect.  5, we compute the linearized AdS supergravity equations
and match the vertex operator field components to the
supergravity fields of general relativity.

\vfill\eject

\newsec{String Constraints}

\bigskip\noindent{\it Preliminaries on $PSU(2|2)$}

The bosonic part of the Lie algebra of $PSU(2|2)$ is the $SU(2)\times SU(2)$
or $SO(4)$ 
Lie algebra.  \foot{For $\AdS_3\times \S^3$, we really
want the real form to be $SU(2)\times SL(2,\R)$, 
 but for our formal discussion we will just take
the real form to be $SU(2)\times SU(2)$.}  The generators
$t^{ab}=-t^{ba}$ of $SU(2)\times SU(2)$ transform as an antisymmetric rank
two tensor of $SU(2)\times SU(2)=SO(4)$.  (Here $a,b=1,\dots,4$ are
vector indices of $SO(4)$.) 
The fermionic generators $e^a$ and $f^a$ transform
as vectors of $SO(4)$.
The Lie superalgebra takes the form
\eqn\gen{\eqalign{ &
[ t_{ab}, t_{cd} ] = \delta_{ac}\,t_{bd} + \delta_{bd}\,t_{ac}
- \delta_{ad}\,t_{bc} - \delta_{bc}\,t_{ad}\,\cr
[ t_{ab}, \, f_c ] = \delta_{ac} f_b - &\delta_{bc} f_a\,,\quad
[ t_{ab}, \, e_c ] = \delta_{ac} e_b - \delta_{bc} e_a\,,\quad
\{ e_a,\, f_b \} = {\half} \epsilon_{abcd} \, t^{cd}\cr
& ~~~~~~~\{e_a,e_b\}=\{f_a,f_b\}=0\,.\cr}}

On the $PSU(2|2)$ manifold, we introduce bosonic and fermionic
coordinates $x^p,\,p=1\dots 6$ and $\theta^a$, $\bar \theta^a$,
$a=1\dots 4$.
We write a generic $PSU(2|2)$ element as
\eqn\groupem{g=\exp({\theta^af_a})h \exp({\bar\theta^{\bar b}e_{\bar b}}),}
with $h\in SU(2)\times SU(2)$.  When we want an explicit parametrization
for $h$, we write
\eqn\turgo{h=\exp(\half\sigma^{pcd}x_pt_{cd}),}
where the $\sigma^{pcd}$ are described in appendix A.  The idea
here is that in the almost flat limit in which $\AdS_3\times \S^3$
reduces to Minkowski space $\R^6$, the $x^p$ become standard Minkowski
coordinates.  In this limit, the $SU(2)\times SU(2)$ is extended
to $SO(6)=SU(4)$, with the $p$ index of $x^p$ transforming as a vector
of $SO(6)$, and indices  $a,b,c$ transforming as a positive
chirality spinor of $SO(6)$ ($\bar a$, $\bar b$, $\bar c$
transform as an $SO(6)$ spinor of the same chirality since we compactify
the Type IIB superstring on ${\bf AdS}_3\times {\bf S}^3\times M$).
We will write $E_a$, $F_a$, and $K_{ab}$ for the operators
that represent the left action of $e_a, $ $f_a$, and $t_{ab}$ on
$g$.  
Very concretely, 
in the above coordinates,
\eqn\ssg{\eqalign{F_a &= {d\over d\t^a}\,,\qquad
K_{ab} = -\t_a {d\over d\t^b} + \t_b {d\over d\t^a} + t_{Lab}\cr
E_a &= {\textstyle{1\over 2}}\epsilon_{abcd}\,\t^b\,
( t_{L}^{cd} - \t^c {d\over d\t_d}\,) + h_{a\bar b} {d\over d\bar\t_{\bar b}}\,.\cr}}
These formulas are found by asking for $F_ag=f_ag$, $E_ag=e_ag$,
$K_{ab}g=-t_{ab}g$.  Also, we have introduced an operator $t_L$
that generates the left action of $SU(2)\times SU(2)$ on $h$ alone,
without acting on the $\theta$'s.
We will also use an analogous $t_R$ that acts on $h$ alone, on the right.
Explicitly
\eqn\ssa{\eqalign{
t_{Lab} \, g \,& = \, e^{\t^a f_a}\, (-t_{ab}) 
\, h(x) \, e^{\bar\t^{\bar a} e_{\bar a}}\,,\quad
t_{R\bar a\bar b} \, g \, = \, e^{\t^a f_a}\,
\, h(x) \, t_{\bar a\bar b}\, e^{\bar\t^{\bar a} e_{\bar a}}\cr}}
where \eqn\g{g = g (x,\t\,, \bar \t)\,
= e^{\t^a f_a}\, 
e^{\hhalf\sigma^{p cd} x_p t_{cd}}\, e^{\bar\t^{\bar a} e_{\bar a}}
= e^{\t^a f_a}\, h(x) \, e^{\bar\t^{\bar a} e_{\bar a}}} 
 In other words, in the parametrization of
$PSU(2|2)$ in terms of $h$, $\theta$, and $\bar\theta$, $t_L$ and $t_R$
generate the left and right $SU(2)\times SU(2)$ action on $h$, leaving
$\theta$ and $\bar\theta$ fixed. 

The use of barred and unbarred indices in the above formulas
is to be understood as follows.  Like any group manifold, the
$SO(4)$ manifold admits a left-invariant vielbein and also a right-invariant
vielbein.  If we write $m,n,p$ for a tangent space index, $A,B,C$
for a local Lorentz index in the right-invariant vielbein, and $\bar A,\bar B,
\bar C$ for a local Lorentz index in the left-invariant vielbein,
then one might write the right- and left-invariant vielbeins
as $e_m^A$ and ${\bar e}_m^{\bar A}$, respectively.  However, because
at every point on the $SO(4)$ manifold, there is a distinguished $SO(4)$
subgroup of the tangent space group $SO(6)$ (namely the subgroup
that leaves fixed the given point), we use an $SO(4)$ notation:
we write a single local Lorentz index $A$ in the vector of $SO(6)$
(or second rank antisymmetric tensor of $SO(4)$) as $ab$ (with antisymmetry
in $a$ and $b$ understood).  Likewise, $\bar A$ is replaced by $\bar a\bar b$.
The right- and left-invariant vielbeins are then written as 
$-\sigma_m^{ab}$ and $\sigma_m^{\bar a\bar b}$, respectively.  $SO(4)$
indices can be raised or lowered as always with the invariant metric
tensor $\delta_{ab}$ of $SO(4)$.  In short, indices $a,b,c$ are
local Lorentz indices for the right-invariant framing transforming
as spinors of $SO(6)$, or vectors of $SO(4)$; and $\bar a,\bar b,\bar c$
are the same for the left-invariant framing.
In \ssg, $h_{a\bar b}$ is the $SO(4)$ group element in the vector representation.
It carries one index of each kind.
It is orthogonal -- that is, $h_{a\bar a} = h^{-1}_{\bar a a}$ -- and
is acted on very simply by $t_L$ and $t_R$:
\eqn\hinverse{\eqalign{h^{a \bar a}\,  t^{cd}_L \, h^{-1}_{\bar a b}
&= - \delta^{ac}\,\delta^{d}_b + \delta^{ad}\,\delta^c_b\,,\qquad
h^{-1}_{\bar a a}\,  t^{\bar c\bar d}_R \, h^{a\bar b}
= - \delta_{\bar a}^{\,\bar c}\,\delta^{\bar d \bar b} 
+ \delta_{\bar a}^{\bar d}\,\delta^{\bar c\bar b}\,.\cr}}

We call the generators of the right action of $PSU(2|2)$ on itself
$\bar K_{\bar a\bar b}$, $\bar E_{\bar a}$, and $\bar F_{\bar a}$.
Just as above, we compute
\eqn\ssgbar{\eqalign{\bar E_{\bar a} &= {d\over d\bar\t^{\bar a}}\,,\qquad
\bar K_{\bar a\bar b} = -\bar\t_{\bar a} {d\over d\bar\t^{\bar b}} 
+ \bar\t_b {d\over d\bar\t^{\bar a}} + t_{R\bar a\bar b}\cr
\bar F_{\bar a} &= {\textstyle{1\over 2}}\epsilon_{\bar a\bar b\bar c\bar d}\,
\bar\t^{\bar b}\,
( - t_{R}^{\bar c\bar d} 
+ \t^{\bar c} {d\over d\bar\t_{\bar d}}\,) 
+ h^{-1}_{\bar a b} {d\over d\t_{b}}\cr}}
from
\eqn\ssabar{\eqalign{\bar F_{\bar a} \,g &= g\,f_{\bar a} \,,\quad
\bar E_{\bar a} \,g = g\, e_{\bar a}\,, 
\quad \bar K_{\bar a\bar b} \,g = g\, t_{\bar a\bar b}\,. \cr}}  
Indices with bars are related to those without bars through the
$SO(4)$ group element $h_{a\bar a}$ by
\eqn\indbar{\eqalign {G_a &= h_a^{\,\,\bar b}\, G_{\bar b},\quad
G_{\bar a} = h_{\bar a}^{-1\,b} \, G_b = h^b_{\,\,\bar a} G_b\,.\cr}}

One of the most important objects  in $PSU(2|2)$ group theory
is the quadratic Casimir operator $f_ae_a+{1\over 8}\epsilon_{abcd}
t^{ab}t^{cd}$.  As an operator on the $PSU(2|2)$ manifold, the
quadratic Casimir is a second order differential operator (since the group
generators are first order operators); we will call it the Laplacian.
The Laplacian can be written in terms of either left or right
generators as 
\eqn\lapl{\eqalign{F_a \, E_a \,
+ {\textstyle{1\over 8}}\e_{a b c d} \, K^{ab}\, K^{cd}\,
&= \bar F_{\bar a} \, \bar E_{\bar a} \,
+ {\textstyle{1\over 8}}\bar\e_{\bar a \bar b \bar c \bar d} \,
\bar K^{\bar a \bar b}\, K^{\bar c\bar d} \cr
=\,h_{a\bar b} \,{d\over d\t_a}\,{d\over d\bar\t_{\bar b}} 
+ {\textstyle{1\over 8}}\e_{a b c d} \, t_L^{ab}\, &t_L^{cd}\,\,
=\,h^{-1}_{\bar b a} \,{d\over d\t_a}\,{d\over d\bar\t_{\bar b}} 
+  {\textstyle{1\over 8}}\bar\e_{\bar a \bar b \bar c \bar d} \,
\bar t_R^{\bar a \bar b}\, t_R^{\bar c\bar d}.\, \cr}} 
In verifying these formulas, one uses the fact that the
 $SO(4)$ Laplacian can similarly be written in terms of left
or right generators:
${\textstyle{1\over 8}}\e_{a b c d} \, t_L^{ab}\, t_L^{cd}
\, = \, {\textstyle{1\over 8}}\bar\e_{\bar a \bar b \bar c \bar d} \,
\bar t_R^{\bar a \bar b}\, t_R^{\bar c\bar d}$. 
It follows from the action of the generators given in \ssgbar$\,$
that $\bar F_{\bar a} , \bar E_{\bar a}, 
\bar K_{\bar a\bar b}$ also obey the $PSU(2|2)$ commutation relations
and that the Laplacian \lapl $\,$ commutes with the group generators 
\ssg, \ssgbar.

We recall finally that $PSU(2|2)$ has an $SL(2,\R)$ group of outer
automorphisms with $E,F$ transforming as a doublet.  In particular,
the substitution
\eqn\var{\eqalign{{{\bar F}'}_{\bar a} = \bar E_{\bar a}\,,\quad
{{\bar E}'}_{\bar a} = - \bar F_{\bar a}\,,\quad
{{\bar K}'}_{\bar a\bar b}  = \bar K_{\bar a\bar b}\cr}}
leaves the Laplacian and the commutation relations invariant. 

\bigskip\noindent{\it Form Of Vertex Operators}

We now want to construct the vertex operators for the supergravity
multiplet.  In this, we are guided by comparison with the
flat space case \refs{\bv,\bvw}.
In the flat case, the supergravity vertex operators are constructed
by  acting with some $N=4$ generators on a spin zero field
$V(x,\theta,\bar\theta;\rho+i\sigma,\bar\rho+i\bar\sigma)$ 
that is a function of these fields but not their derivatives.  $V$
also obeys certain additional constraints.

The spin zero property is ensured by having $V$ annihilated by
the Laplacian $\partial^p\partial_p$. The Laplacian is a
second order differential operator that is invariant under the
Poincar\'e symmetries.  The analogous invariant second order
operator on the $PSU(2|2)$ manifold is the quadratic Casimir operator
that was introduced above.  

The ghost fields
$\rho$ and $\sigma$ enter only in the combinations $\rho+i\sigma$
and its barred counterpart because these combinations have zero
background charge and so do not contribute to the dimension of the operator.
Also, these are the combinations that remain holomorphic after perturbing
to $\AdS_3\times \S^3$ as in \bvw.  

Finally, the constraint equations in the flat space case, as presented
in \refs{\bv,\bvw}, are
written in terms of $\partial^{ab}=\sigma^{p ab}\partial/\partial x^p$
and $\nabla_a=\partial/\partial\theta^a$.  In the curved case,
$\partial^{ab}$ is deformed to $K^{ab}$ and $\nabla_a$ is interpreted 
as $F_a$.  To the flat space constraint equations, we find that
we have to add lower order terms to achieve $PSU(2|2)$ invariance.

The expansion of the vertex operator in terms of the ghost fields is
\eqn\vee{V = \sum_{m,n = -\infty}^\infty\,
e^{m(i\s + \rho) + n(i\bar\s + \bar\rho)}\,
V_{m,n} (x, \t, \bar\t)\,.}
In flat space, the constraints from the left and right-moving 
worldsheet super Virasoro algebras are from \bvw:
\eqn\cononef{\eqalign{(\N)^4 V_{1,n} &= \N_a \,\N_b
\p^{a b} V_{1,n} = 0\cr
{\textstyle{1\over 6}}\e^{a b c d} \,\N_b \,\N_c \,\N_d
V_{1,n} &= - i \N_b\, \p^{a b} V_{0,n}\cr
\N_a\,\N_b\, V_{0,n} - {\textstyle{i\over 2}} \e_{a b c d} \, \p^{cd}\,
V_{-1,n} &= 0\,,\qquad \N_a \, V_{-1,n} = 0\,;\cr}}
\eqn\contwof{\eqalign{\bar\N^4 V_{n,1} &= \bar\N_{\bar a}\bar\N_{\bar b}
\bar\p^{\bar a\bar b} V_{n,1} = 0\cr
{\textstyle{1\over 6}} \e^{\bar a\bar b\bar c\bar d}\bar
\N_{\bar b}\bar \N_{\bar c} \bar \N_{\bar d}
V_{n,1} &= -i \bar \N_{\bar b} \bar\p^{\bar a\bar b} V_{n,0}\cr
\bar\N_{\bar a}\bar \N_{\bar b} V_{n,0}
- {\textstyle{i\over 2}} \bar\e_{\bar a \bar b \bar c \bar d}
\, \bar\p^{\bar c\bar d}\, V_{n,-1} &= 0\,,
\qquad \bar\N_{\bar a} \, V_{n,-1} = 0\cr}}
\eqn\conthreef{\eqalign{\p^p\p_p V_{m,n} = 0\cr}}
for $-1\le m,n\le 1$, with the notation
$\N_a = d/ d\t^a$, $\bar\N_{\bar a} = d/ d\bar\t^{\bar a}$,
$\partial^{ab} = -\sigma^{p ab}\,\p_p$.
In flat space, these equations were derived by requiring the
vertex operators to satisfy the physical state conditions
\eqn\svaf{\eqalign{G^-_0 V=\tilde G^-_0 V=\bar G^-_0 V=
\bar{\tilde G}^-_0 V =T_0 V=\bar T_0 V= 0,\cr
J_0 V = \bar J_0 V = 0\,,\quad G^+_0\tilde G^+_0 V=
\bar G^+_0\bar{\tilde G}^+_0 V=0\cr}}
where $T_n,\, G^\pm_n,\, \tilde  G^\pm_n,\,J_n, J^\pm_n$ and
corresponding barred generators
are the left and right \break $N=4$ worldsheet superconformal generators
present in the Berkovits-Vafa quantization
of the type II superstring \refs{\bv,\b, \bvw}.
The conditions \svaf~further implied
$V_{m,n} = 0$ for $m>1$ or $n>1$ or $m<1$ or $n<1$, leaving
nine non-zero components.

In curved space, we modify these equations as follows:
\eqn\conone{\eqalign{F^4 V_{1,n} &= F_a \,F_b
K^{a b} V_{1,n} = 0\cr
{\textstyle{1\over 6}}\e^{a b c d} \,F_b \,F_c \,F_d
V_{1,n} &= - i F_b\, K^{a b} V_{0,n} +  2i F^a  V_{0,n} - E^a V_{-1,n}\cr
F_a\,F_b\, V_{0,n} - {\textstyle{i\over 2}} \e_{a b c d} \, K^{cd}\,
V_{-1,n} &= 0\,,\qquad F_a \, V_{-1,n} = 0\,;\cr}}
\eqn\contwo{\eqalign{\bar F^4 V_{n,1} &= \bar F_{\bar a}\bar F_{\bar b}
\bar K^{\bar a\bar b} V_{n,1} = 0\cr
{\textstyle{1\over 6}} \e^{\bar a\bar b\bar c\bar d}\bar 
F_{\bar b}\bar F_{\bar c} \bar F_{\bar d}
V_{n,1} &= -i \bar F_{\bar b} \bar K^{\bar a\bar b} V_{n,0}
+ 2i \bar F^{\bar a}  V_{n,0} - \bar E^{\bar a} V_{n,-1}\cr
\bar F_{\bar a}\bar F_{\bar b} V_{n,0} 
- {\textstyle{i\over 2}} \bar\e_{\bar a \bar b \bar c \bar d} 
\, \bar K^{\bar c\bar d}\, V_{n,-1} &= 0\,,
\qquad \bar F_{\bar a} \, V_{n,-1} = 0.\cr}}
There is also a spin zero condition constructed from the Laplacian:
\eqn\conthree{\eqalign{(\,
F_a \, E_a \,+ {\textstyle{1\over 8}}\e_{a b c d} \, K^{ab}\, K^{cd}\,)
\, V_{n,m} &=  
(\,\bar F_{\bar a} \, \bar E_{\bar a} \,
+ {\textstyle{1\over 8}}\bar\e_{\bar a \bar b \bar c \bar d} \, 
\bar K^{\bar a \bar b}\, K^{\bar c\bar d}\,)
\, V_{n,m} = 0\,.\cr}} 
Equations \conone -\conthree $\,$  
have been derived by deforming the corresponding equations
for the flat case, which were presented above,
by requiring invariance  under the $PSU(2|2)$
transformations that we will present momentarily.  
For example, in the second equation
in \conone, the terms on the right hand side that are linear
in $PSU(2|2)$ generators are curvature corrections to the flat
equations, while the quadratic and cubic terms are present in the flat
case. 

The above constraint equations are invariant under the action of
$PSU(2|2)\times PSU(2|2)$, but there is a subtlety in how $PSU(2|2)$ acts.
The subtlety comes from the unusual form of the $PSU(2|2)\times PSU(2|2)$
currents found in \bvw\ (and ultimately from the unusual form of
half of the supercharges even in the flat case in this formalism \bv).
While $F$ and $K$ have natural definitions in \bvw, and act by the
obvious left multiplication on $g$ while leaving invariant the ghosts,
this is not true for $E$.  Rather, $E$ has a term that acts naturally
on $g$, plus corrections that are proportional to $e^{-\rho-i\sigma}$
times $F$.

We have found that the following transformations generate a $PSU(2|2)$
action on the $V$'s that commutes with the constraint equations
in \conone - \conthree:
\eqn\sym{\eqalign{\Delta_a^-\, V_{m,n} &= F_a \, V_{m,n}\,,\quad
\Delta_{ab}\, V_{m,n} = K_{ab} \, V_{m,n}\cr
\Delta_a^+\, V_{1,n} = E_a \, V_{1,n}\,,\quad
\Delta_a^+\, V_{0,n} &= E_a \, V_{0,n} + i F_a V_{1,n}\,,\quad 
\Delta_a^+\, V_{-1,n} = E_a \, V_{-1,n} - i F_a V_{0,n}\,.\cr}}
(In a hopefully obvious notation, we write $\Delta_{ab}$ for the variation
of the vertex operator generated by $t_{ab}$ and $\Delta_a^{\pm}$ for
the variations generated by $e_a$ and $f_a$.)
These formulas are the standard $PSU(2|2)$ generators on the $V_{m,n}$,
except that to $E$ we have added a multiple of $F$ times a ``raising''
operator on the $m$ index.  This is suggested by the form of the
supercharges $q_a^\pm$ in \bvw.  
The constraint equations in \conone -\conthree\
were determined by being invariant under the deformed $PSU(2|2)$ generators.
This uniquely fixed all the ``lower order terms'' that were not present
in the flat case.  It would be desireable to derive \conone -\conthree\
directly by constructing the constraint operators and their action
on vertex functions exactly, or at least perturbatively in $\alpha'$.
However, the uniqueness gives us confidence in the formulas.

\newsec{Gauge Transformations}

The constraint equations for the vertex operators
are also invariant under gauge transformations. We use these
to identify the physical degrees of freedom of the vertex operator
\vee~ as follows. 
Useful gauge symmetries of the flat space equations are
\eqn\gaugef{\eqalign{\delta V &=
G_0^+ \Lambda + \tilde G_0^+ \tilde\Lambda 
+ \bar G_0^+ \bar\Lambda + \bar {\tilde G_0}^+ \bar{\tilde\Lambda}\,,
\qquad \delta V = \tilde G_0^+  \tilde G_0^-
\bar{ \tilde G_0^+} \bar{\tilde G_0^-} \Omega\cr}}
with (sum on $n =-1,0,1$)
$$\eqalign{\Lambda &= 
e^{i\s + 2\rho  + n (i\bar\s + \bar\rho ) }\,\lambda_n(x,\t ,\bar\t)\,,
\qquad \bar\Lambda =
e^{i\bar\s + 2\bar\rho  + n (i\s + \rho ) }\,
\bar\lambda_n(x,\t ,\bar\t)\cr
\tilde\Lambda &=
e^{-\rho  - i H_c + n (i\bar\s + \bar\rho ) }\,
\tilde\lambda_n(x,\t ,\bar\t)\,,
\qquad \bar{\tilde\Lambda} =
e^{-\bar\rho -i\bar H_c + n (i\s + \rho ) }\,
\bar{\tilde\lambda}_n(x,\t ,\bar\t)\cr}$$
where the gauge parameters $\Lambda$,$\tilde\Lambda$ are annihilated by
$\bar G_0^+ \bar{\tilde G}_0^+$, the gauge parameters
$\bar\Lambda$,$\bar{\tilde\Lambda}$ are annihilated by
$G_0^+ {\tilde G}_0^+$, all $\Lambda, \tilde\Lambda,
\bar\Lambda, \bar{\tilde\Lambda}$ are annihilated by\nobreak
$\,T_0, \bar T_0, G_0^-, \tilde  G_0^-, \bar  G_0^-, \bar{\tilde  G}_0^-$,
and $\Omega$ is annihilated by $T_0,\bar T_0$. 
The currents $e^{-i H_c}, e^{-i \bar H_c}$ are related to the $N=4$ currents
$J^-(z) = e^{-\rho - i\s} e^{-i H_c}$, 
$\bar J^-(\bar z) = e^{-\bar\rho - i \bar\s} e^{-i \bar H_c}$. 
Since fermionic conformal fields do not commute, {\it i.e.} 
$e^{i\s(z)} \, e^{\rho(\zeta)} \sim - e^{\rho(\zeta)} \, e^{i\s(z)}$,
the notation is defined as
$e^{i\s + \rho} \equiv e^{i\s}\, e^\rho$, etc.

Using the transformations \gaugef~
one can gauge fix to zero the vertex operators
$V_{-1,1}, V_{1,-1}, V_{0,-1}, V_{-1,0}, V_{-1,-1}$, and therefore
they do not correspond to propagating degrees of freedom. 
Furthermore this gauge symmetry can be used both to set to zero
the components of $V_{1,1}$ with no $\t$'s or no $\bar\t$'s,
and to gauge fix all components of $V_{0,1}, V_{1,0}, V_{0,0}$ 
that are independent of those of $V_{1,1}$.
The physical degrees of freedom are thus 
described by a superfield \refs{\bvw}  
\eqn\vone{\eqalign{V_{1,1}&=\t^a\bar\t^{\bar a} V^{--}_{a\bar a} +
\t^a\t^b\bar\t^{\bar a}\s^m_{ab} \bar\xi^-_{m\,\bar a}+
\t^a\bar\t^{\bar a} \bar\t^{\bar b}\s^m_{\bar a\bar b} 
\xi^-_{m\, a}\cr
&+\t^a\t^b\bar\t^{\bar a} \bar\t^{\bar b}\s^m_{ab}\s^n_{\bar a\bar b}
( g_{mn}+b_{mn}+\bar g_{mn}\phi) +
\t^a(\bar\t^3)_{\bar a} A^{-+\,\bar a}_{a}
+(\t^3)_a\bar\t^{\bar a} A^{+-\, a}_{\bar a}\cr
&+\t^a\t^b(\bar\t^3)_{\bar a}\s^m_{ab} \bar\chi_m^{+\,\bar a}+
(\t^3)^a\bar\t^{\bar a} \bar\t^{\bar b}
\s^m_{\bar a\bar b} \chi_m^{+\, a}+
(\t^3)_a(\bar\t^3)_{\bar a} F^{++\, a\bar a}\,.\cr}} 
This has the field content of $D=6$, $N=(2,0)$ supergravity
with one supergravity and one tensor multiplet \refs{\r}. 
As explained in \refs{\bvw}, in flat space, 
the surviving constraint equations in the set \cononef -\conthreef~
imply that the component fields $\Phi$ are all on shell massless
fields, that is
$\sum_{m=1}^6\p^m \p_m \Phi=0$ and in addition
\eqn\sgf{\eqalign{\p^m g_{mn} &= -\p_n\phi\,,\quad  \p^m b_{mn} = 0\,,\quad
\p^m \chi^{\pm b}_m = \p^m \bar\chi^{\pm\bar b}_m = 0\cr
\partial_{ab} \chi_m^{\pm b} &= \partial_{\bar a\bar b} 
\chi_m^{\pm \bar b} = 0\,,\quad 
\partial_{cb} F^{\pm\pm b\bar a} =
\partial_{\bar c\bar b} F^{\pm\pm \bar b a} = 0\,,\cr}}
where 
$$\eqalign{F^{+- a\bar a} = \p^{\bar a\bar b}
A_{\bar b}^{+- a}&\,,\quad
F^{-+ a\bar a} = \p^{a b}
A_b^{-+ \bar a}\,,\quad 
F^{-- a\bar a} = \p^{a b} \p^{\bar a\bar b} V_{b\bar b}^{--}\cr
&\chi_m^{-a} = \p^{ab} \xi_{m b}^-\,, \quad
\bar\chi_m^{- \bar a} = \p^{\bar a\bar b} \bar\xi_{m{\bar b}}^-\,.\cr}$$
The equations of motion \sgf~ for the flat space vertex operator component fields
describe $D=6$, $N=(2,0)$ supergravity 
expanded around the six-dimensional Minkowski metric. 

In $\AdS_3\times \S^3$ space there are corresponding gauge transformations
which reduce the number of degrees of freedom to those in \vone~,
but the Laplacian must be replaced by the AdS Laplacian, and
the constraints are likewise deformed.
In section 4, we focus on the vertex operator $V_{11}$ that carries
the physical degrees of freedom. 
We derive the conditions on its field components
that follow from the AdS vertex operator constraint equations 
\conone-\conthree , and give their residual gauge invariances explicitly.
In section 5, we show these conditions are equivalent to 
the $D=6, N=(2,0)$ linearized supergravity equations expanded around
the $\AdS_3\times \S^3$ metric.

\newsec{String Equations for AdS Vertex Operator Field Components}

Acting on the superfield in \vone, the 
AdS supersymmetric constraints \conone\ - \conthree\ imply
\eqn\confour{\eqalign{
F_a \,F_b \,K^{a b} V_{1,1} &= 0\,,\qquad
\bar F_{\bar a}\bar F_{\bar b}\,\bar K^{\bar a\bar b} V_{1,1} = 0\cr}}
\eqn\confive{\eqalign{
(\, F_a \, E_a \,+ {\textstyle{1\over 8}}\e_{a b c d} \, K^{ab}\, K^{cd}\,)
\, V_{1,1} &=
(\,\bar F_{\bar a} \, \bar E_{\bar a} \,
+ {\textstyle{1\over 8}}\bar\e_{\bar a \bar b \bar c \bar d} \,
\bar K^{\bar a \bar b}\, K^{\bar c\bar d}\,)
\, V_{1,1} = 0\,.\cr}}   
For the bosonic field components of the vertex operators, 
the zero Laplacian condition \confive~ requires that 
\eqn\ccone{\eqalign{
&\Box \,h^g_{\,\,\bar a} \,V^{--}_{ag} =
-4 \,\sigma^m_{ab}\,\sigma^n_{gh}\,\delta^{bh}\, h^g_{\,\,\bar a}\,
G_{mn}\cr}}
\eqn\cctwo{\eqalign{&\Box \, h^g_{\,\,\bar a}\,
h^h_{\,\,\bar b}\, \sigma^m_{ab}\,\sigma^n_{gh}\, G_{mn} =
{\textstyle{1\over 4}} \epsilon_{abce} \epsilon_{fghk}
\, \delta^{ch}\,h^f_{\,\,\bar a}\,  h^g_{\,\,\bar b}\,
\, F^{++ e k}\cr}}
\eqn\ccthree{\eqalign{&\Box \, h_g^{\,\,\bar a}\, 
F^{++ a g} = 0\,,\quad
\Box \, h_g^{\,\,\bar a}\, A_a^{-+  g} = 0\,,\quad 
\Box \, h^g_{\,\,\bar a}\, A_g^{+-  a} = 0\cr}}
and \confour~ results in 
\eqn\ccfour{\eqalign
{\epsilon_{eacd}\, t^{cd}_L\, h^b_{\,\,\bar a}\, A_b^{+-a} = 0\,,\qquad
\epsilon_{\bar e\bar b\bar c\bar d}\, t^{\bar c\bar d}_R\,
h_a^{\,\,\bar a}  \,A_{\bar a}^{-+ \bar b} = 0\cr}}
\eqn\ccfive{\eqalign{
\epsilon_{eacd}\, t^{cd}_L\, h_b^{\,\,\bar a}\, F^{++ab} = 0\,,\qquad
\epsilon_{\bar e\bar b\bar c\bar d}\, t^{\bar c\bar d}_R\,
h^a_{\,\,\bar a}  \,F^{++\bar a \bar b} = 0\cr}}
\eqn\ccsix{\eqalign{ 
&t^{ab}_L\, h^g_{\,\,\bar a}\,  h^h_{\,\,\bar b}\,\sigma^m_{ab}\,\sigma^n_{gh}\,
G_{mn} =\,0\,,\quad
t^{\bar a\bar b}_R\, h^{\,\,\bar g}_a\,  h^{\,\,\bar h}_{b}\,
\sigma^m_{\bar g\bar h}\,\sigma^n_{\bar a \bar b}\, G_{mn}\,
=\,0\,.\cr}}
We have used \indbar~ to relate barred and unbarred indices. 
We have expanded $G_{mn}= g_{mn} + b_{mn} + \bar g_{mn}\phi$.
The $SO(4)$ Laplacian is
$\Box\equiv\nobreak{\textstyle{1\over 8}} \epsilon_{abcd} \,t^{ab}_L\, t^{cd}_L
\,=\, {\textstyle{1\over 8}} \epsilon_{\bar a\bar b\bar c\bar d}
\,t^{\bar a\bar b}_R\, t^{\bar c \bar d}_R$.  
{}From identities such as 
$\half \e_{deab} \, (\,t_L^{cd}\, t_L^{ab}\, +
\,t_L^{ab}\, t_L^{cd}\,) = - {\textstyle{1\over 4}}\,
\delta_e^c\, \e_{abfg}\ t_L^{ab}\, t_L^{fg}$
we find \ccthree~ follows from \ccfour,\ccfive.

In order to compare this with supergravity, we want to reexpress
the above formulas in terms of covariant derivatives $D_p$ on
the group manifold.  We will, however, write everything in terms of
right- or left-invariant vielbeins described in section 2 and appendix B.
So we write 
\eqn\covdd{\eqalign{{\cal T}_L^{cd} &\equiv -\sigma^{p\,cd} \, D_p\,,\qquad
{\cal T}_R^{\bar c\bar d} \equiv \sigma^{p\, \bar c\bar d} \, D_p\,\,.\cr}}
We also have corresponding objects $t_L^{cd}=-\sigma^{p\,cd}D'_p$,
$t_R^{cd}=\sigma^{p\,\bar c\bar d}D''_p$, where (in contrast to $D_p$
which is the covariant derivative with the Levi-Civita connection)
$D'$ and $D''$ are covariant derivatives defined such that right- or
left-invariant vector fields are covariantly constant.
Acting on a function, ${\cal T}_L=t_L$ and ${\cal T}_R=t_R$, since
both just act geometrically.
But they differ in acting on fields that carry spinor or vector indices.
For example, on spinor indices, 
\eqn\oneinvdt{\eqalign{t_L^{ab} V_e&=
{\cal T}_L^{ab} \, V_e + \half \delta^a_e\, \delta^{bc} V_c
- \half \delta^b_e\, \delta^{ac} V_c\,\cr}} 
so that the invariant derivatives on the $SO(4)$  group manifold satisfy 
\eqn\invd{\eqalign{t_L^{cd} \, \sigma ^n_{ef} \, f_n(x) &\equiv
{\cal T}_L^{cd} \, \sigma ^n_{ef} \,f_n(x) \, + \, {\textstyle{1\over 2}}
f_{ef}^{\,\,\, ghcd}\, \sigma ^n_{gh} \, f_n(x)\cr
t_R^{\bar c\bar d} \, \sigma ^n_{\bar e\bar f}  \, f_n(x)&\equiv
{\cal T}_R^{\bar c \bar d} \, \sigma ^n_{\bar e\bar f}  \, f_n(x)\, 
- \, {\textstyle{1\over 2}}
f^{\bar c\bar d \bar g\bar h}_{\qquad \bar e\bar f} \sigma ^n_{\bar g\bar h}
\, f_n(x)\cr}}
where
the $SO(4)$ structure constants from \gen~ are
\eqn\strc{\eqalign{f_{ef}^{\,\,\, ghcd} = {\textstyle{1\over 2}} 
& [\, \delta^{gc}\delta_e^h\delta_f^d - \delta^{gd}\delta_e^h\delta_f^c 
- \delta^{hc}\delta_e^g\delta_f^d + \delta^{hd}\delta_e^g\delta_f^c \cr
& - (e \leftrightarrow f)\,]\,.\cr}}
These definitions of the invariant derivatives are, for
example, compatible with
\eqn\jipog{{\textstyle{1\over 8}} \epsilon_{abcd} \,t^{ab}_L\, t^{cd}_L
\,\, \,\sigma ^n_{ef}\,f_n(x) =
\, {\textstyle{1\over 8}} \epsilon_{\bar a\bar b\bar c\bar d}
\,t^{\bar a\bar b}_R\, t^{\bar c \bar d}_R \,\,\,\sigma ^n_{ef}\,f_n(x).}
For a more detailed discussion, see appendix B.

\vfill\eject
\subsec{ Gauge Conditions}

We show that the constraints on the vertex operators \ccone-\ccsix~
are identical to those of AdS supergravity as follows. 
Using 
$$\eqalign{t^{ab}_L\, h^g_{\,\,\bar a}\,  h^h_{\,\,\bar b}\,\sigma^m_{ab}
\,\sigma^n_{gh}\,G_{mn} =&
(t^{ab}_L\, h^g_{\,\,\bar a}\,)  h^h_{\,\,\bar b}\,\sigma^m_{ab}
\,\sigma^n_{gh}\,G_{mn}\cr
&+h^g_{\,\,\bar a}\, (t^{ab}_L\,  h^h_{\,\,\bar b}\,) \sigma^m_{ab}
\,\sigma^n_{gh}\,G_{mn} +
 h^g_{\,\,\bar a}\,  h^h_{\,\,\bar b}\,(t^{ab}_L\, \sigma^m_{ab}
\,\sigma^n_{gh}\,G_{mn})}$$
and \hinverse~, we find that \ccsix~ are gauge conditions on the string 
fields:
\eqn\gc{\eqalign{D^p g_{ps} &= - D_s\phi -{\textstyle{1\over 2}}
\,(\sigma^m \sigma_s \sigma^n)_{ab}\,\delta^{ab}\, b_{mn} = D^p g_{sp}\,,
\qquad
D^p b_{ps} = 0 = D^p b_{sp}.\,\cr}}
This is the curved space analog of the flat space conditions
$\p^p g_{ps} = -\p_s\phi\,,\,\,\p^p b_{ps} = 0$. 
The string equations \ccone-\ccsix\ are invariant 
under the residual gauge transformations which transform $G_{mn}$:
\eqn\resg{\eqalign{ \bigtriangleup g_{ps} =\, &
{\textstyle{1\over 2}} ( D_p\xi_s + D_s\xi_p ) \,,\cr
\bigtriangleup b_{ps} = \,&
{\textstyle{1\over 2}} ( D_p\eta_s - D_s\eta_p ) + {\textstyle{1\over 2}}
(\sigma_p \sigma_s \sigma_q)_{ab}\delta^{ab} \xi^q \,,\cr
\bigtriangleup \phi \,=\, & 0\,,\cr
\bigtriangleup V_{ag}^{--} \,=\, &\half \e_{abcd}\, 
t_L^{cd}\Lambda^b_{\,\,g}\,  + \half  h_g^{\,\,\bar a}
\e_{\bar a\bar b\bar c\bar d}\, 
t_R^{\bar c\bar d}\bar\Lambda_a^{\,\,\bar b}\cr}} 
where the gauge parameters $\xi_m,\eta_m$ satisfy
$$\eqalignno{&D^p D_p \xi_m + (\sigma^p \sigma_m \sigma^n)_{ab}\delta^{ab}
D_p \eta_n + \bar R_{mp}\xi^p = 0\,,\cr
&D^p D_p \eta_m + (\sigma^p \sigma_m \sigma^n)_{ab}\delta^{ab}
D_p \xi_n + \bar R_{mp}\eta^p = 0\,,\cr
&D^m\xi_m = D^m\eta_m =\,0\,,\cr}$$
$V^{--}_{ag}$ transforms with gauge parameters
$$\eqalignno{&\Box \Lambda^b_{\,\,\bar a} =
4 h^{b\bar b} \s^m_{\bar b\bar a} (\xi_m + \eta_m)\,,
\qquad \Box \bar\Lambda_a^{\,\,\bar b} =
- 4 h^{b\bar b}  \s^m_{ab} \,(\xi_m - \eta_m)\,\cr}$$
and the $\AdS_3\times \S^3$ Ricci tensor $\bar R_{rp}$ is 
\eqn\Ricci{\eqalign{\bar R_{rp} &\equiv - {\textstyle{1\over 2}} \,
\sigma_r^{ab} \sigma_p^{cd} \delta_{ac} \delta_{bd}\,.\cr}} 
The other string fields are invariant. There are additional 
independent gauge symmetries  
which transform $A^{+-b}_{\bar a}, A^{-+\bar b}_a$, but leave invariant
$F^{+-a\bar b}, F^{-+a\bar b}$ defined in section 4.3.
For $\AdS_3\times \S^3$  we can write the Riemann tensor
and the metric tensor as
\eqn\Riemann{\eqalign{\bar R_{mnp\tau} &=
{\textstyle{1\over 4}} \, (\, \bar g_{m\tau} \bar R_{np} +
\bar g_{np} \bar R_{m\tau} - \bar g_{n\tau } \bar R_{mp} -
\bar g_{mp} \bar R_{n\tau}\,)\cr
\bar g_{mn} &=
{\textstyle{1\over 2}}\, \sigma_m^{ab}\,\sigma_{n\,ab}\,.\cr}}

\subsec{ Metric, Dilaton, and Two-form}

We find from the string constraint \cctwo~ that the six-dimensional
metric field $g_{rs}$,
the dilaton $\phi$, and the two-form $b_{rs}$ satisfy
\eqn\ssone{\eqalign{{\textstyle{1\over 2}} D^p D_p b_{rs} =&
-{\textstyle{1\over 2}} (\sigma_r\sigma^p\sigma^q)_{ab}\delta^{ab}
\,D_p\,
[\,g_{qs} + \bar g_{qs}\phi\,]
+{\textstyle{1\over 2}} (\sigma_s\sigma^p\sigma^q)_{ab}\delta^{ab}
D_p\,
[\,g_{qr} + \bar g_{qr}\phi\,]\cr
&-\bar R_{\tau r s \lambda} \, b^{\tau\lambda}\,
-{\textstyle{1\over 2}} \bar R_r^{\,\,\tau}\, b_{\tau s}
-{\textstyle{1\over 2}} \bar R_s^{\,\,\tau}\, b_{r\tau}\cr
&+ {\textstyle{1\over 4}} F^{++gh}_{\rm asy}
\,\sigma_r^{ab}\sigma_s^{ef}\, \delta_{ah}\delta_{be}\delta_{gf}
\cr}}
\eqn\sstwo{\eqalign{{\textstyle{1\over 2}} D^p D_p \,
(\, g_{rs} + \bar g_{rs} \phi \,)\,=
&-{\textstyle{1\over 2}} (\sigma_r\sigma^p\sigma^q)_{ab}\delta^{ab}    
\,D_p b_{qs} \,
+ {\textstyle{1\over 2}} (\sigma_s\sigma^p\sigma^q)_{ab}\delta^{ab}
\,D_p b_{rq} \cr
&- \bar R_{\tau r s \lambda} \,
(\, g^{\tau\lambda} + \bar g^{\tau\lambda}\phi \,)\,
-{\textstyle{1\over 2}} \bar R_r^{\,\,\tau}\,
(\, g_{\tau s} + \bar g_{\tau s} \phi \,)
-{\textstyle{1\over 2}} \bar R_s^{\,\,\tau}\,
(\, g_{r\tau } + \bar g_{r\tau }\phi \,)\cr
&+ {\textstyle{1\over 4}} F^{++gh}_{\rm sym} \,\sigma_{rga}\sigma_{shb}\,
\delta^{ab}\,.\cr}}
This is the curved space version of the flat space zero Laplacian condition
$\p^p \p_p b_{rs} = \p^p \p_p g_{rs} = \p^p \p_p \phi = 0$ from \sgf. 

\subsec{Self-Dual Tensors and Scalars}

Four self-dual tensor and scalar pairs come from the string bispinor
fields\break 
$ F^{++ ab}, \,V^{--}_{ab}, \,,A^{+- a}_b, \,A^{-+ b}_a$.
{}From \ccfive, we have
\eqn\ssthree{\eqalign
{&\sigma^p_{da}\, D_p \, F^{++ab}_{\rm asy} = 0\cr}}
\eqn\ssfour
{\eqalign{{\textstyle{1\over 4}}\,[\, \delta^{Ba} \,\sigma^r_{ga}\,D_r\,
F^{++gH}_{\rm sym} \, - \,
\delta^{Ha} \,\sigma^r_{ga}\,D_r\, F^{++gB}_{\rm sym} \, ]
= -{\textstyle{1\over 4}}\,\, \epsilon^{BH}_{\qquad cd}\,\,
F^{++cd}_{\rm asy}\,.\cr}} 
{}From \ccone, we find
\eqn\ssfive{\eqalign{&{\textstyle{1\over 2}}\, D^p D_p\, V_{cd}^{--}
\, -{\textstyle{1\over 2}} \delta^{gh}\sigma^p_{ch}\, D_p \, V_{gd}^{--}
\, +{\textstyle{1\over 2}} \delta^{gh}\sigma^p_{dh}\, D_p \, V_{cg}^{--}
\, +{\textstyle{1\over 4}} \epsilon_{cd}^{\,\,\,\,gh}\, V_{gh}^{--}\cr
=&\,-4\,\sigma^m_{ce}\,\sigma^n_{df}\,\delta^{ef}\, G_{mn}\,.\cr}}  
The string constraints \ccfour~ can be written as
\eqn\sssix{\eqalign{&\epsilon_{eacd}\,
t^{cd}_L\, h_b^{\,\,\bar a}\, F^{+-ab} = 0\,\qquad
\epsilon_{\bar e\bar b\bar c\bar d}\, t^{\bar c\bar d}_R\,
h^a_{\,\,\bar a}  \,F^{+-\bar a \bar b} = 0\cr
&\epsilon_{eacd}\, t^{cd}_L\, h_b^{\,\,\bar a}\, F^{-+ab} = 0\,,
\qquad\epsilon_{\bar e\bar b\bar c\bar d}\, t^{\bar c\bar d}_R\,
h^a_{\,\,\bar a}  \,F^{-+\bar a \bar b} = 0\cr}}
where
$$\eqalignno{&F^{+-a\bar a}\equiv \delta^{\bar a\bar b} \,
A_{\bar b}^{+- a} + t_R^{\bar a\bar b} \, A_{\bar b}^{+- a}\cr
&F^{-+a\bar a}\equiv \delta^{ab} \,
A_{b}^{-+ \bar a} + t_L^{ab} \, A_b^{-+ \bar a}\,,\cr}$$ 
which is similar to the form of \ccfive~, so $F^{+-ab}$ and $F^{-+ab}$
also satisfy equations of the form \ssthree,\ssfour.

\subsec{Gravitinos and Spinors} 

Independent conditions on the fermion fields from \confour,\confive~ are
\eqn\ccfer{\eqalign{&\Box \, h_a^{\,\,\bar g}\,
\s^m_{\bar a\bar b}\, \xi^-_{m \bar g}
= - \s^m_{\bar g\bar h}\,\e_{\bar e\bar d\bar a\bar b} \,
h_a^{\,\,\bar h}\, \delta^{\bar g\bar d} \,\bar\chi_m^{+\bar e}\cr
&\Box \, h^g_{\,\,\bar a}\,
\s^m_{a b}\, \bar\xi^-_{m g}
= - \s^m_{gh}\,\e_{edab} \,
h^h_{\,\,\bar a}\, \delta^{gd} \,\chi_m^{+ e}\cr
&t_L^{\,\, ab}\, h^g_{\,\bar a}\, \sigma^m_{ab}\,
\bar\xi_{m g}^{-} \, = 0\,,\qquad
t_R^{\,\, \bar a\bar b}\,  h_a^{\,\bar g}\,\sigma^m_{\bar a\bar b}\,
\xi_{m \bar g}^{-} \, = 0\cr
&t_L^{\,\, ab}\, \sigma^m_{ab}\, h_g^{\,\bar a}\,
\bar\chi_m^{+\,g} \, = 0\,,\qquad
t_R^{\,\, \bar a\bar b}\, \sigma^m_{\bar a\bar b}\,
h^a_{\,\bar g}\, \chi_m^{+\,\bar g} \, = 0\cr
&\epsilon_{deab}\, t_L^{\,ab}\,
h^g_{\,\,\bar a}\, h^h_{\,\,\bar b}\,\sigma^m_{gh}\,
\chi_m^{+\,e} \, = 0\,,\qquad
\epsilon_{\bar d\bar e\bar a\bar b}\, t_R^{\,\bar a\bar b}\,
h_a^{\,\,\bar g}\, h_b^{\,\,\bar h}\,\sigma^m_{\bar g\bar h}
\bar\chi_m^{+\,\bar e} \, = 0\,.\cr}}

\newsec{$D=6$, $N=(2,0)$ Supergravity on $\AdS_3\times \S^3$}

In this section we show that the string constraints described in
section 5, which were derived from the $\AdS_3\times \S^3$
supersymmetric vertex operator equations \conone-\conthree,
are equivalent to the supergravity equations
for the supergravity multiplet and one tensor multiplet
of $D=6$, $N=(2,0)$ supergravity expanded around the
${\bf AdS}_3\times {\bf S}^3$ metric and a self-dual three-form. 
We give the identification of the string vertex operator components
in terms of the supergravity fields. 

In flat space, the vertex operator field components can be
described by representations of the $D=6$ little group $SO(4)$.
The supergravity multiplet is $(3,3) + 5 (3,1) + 4 (3,2)$
and the tensor multiplet is $(1,3) + 5 (1,1) + 4 (1,2)$.
Here $(3,3)$ is the graviton field $g_{mn}$, 
the oscillation around the flat metric; the two-form $b_{mn}$ has
SO(4) spin content $(1,3) + (3,1)$; and the dilaton $\phi$ is
$(1,1)$. The bispinor fields
$ F^{++ ab}, \,V^{--}_{ab}, \,A^{+- a}_b, \,A^{-+ b}_a$
represent four self-dual tensors $(3,1)$ and four scalars $(1,1)$.
The four gravitinos $4 (3,2)$ and four spinors $4 (1,2)$
are labelled by $\chi^{+a}_m,\,\bar\chi^{+\bar a}_m,\,
\xi^-_{ma},\,\bar\xi^-_{m\bar a}$.

In $\AdS_3\times \S^3$ space, the 
number of physical degrees of freedom
in the vertex operator \vee~ remains the same as in the 
flat case, but
the field $g_{mn}$ is now related to 
the oscillation around the $\AdS_3\times \S^3$
metric. For this metric to be a solution of the 
supergravity equations \refs{\r}, 
there must be also either a non-vanishing (real) self-dual tensor 
field or
a non-vanishing (real) anti-self-dual tensor field. 
Expanding around this classical solution,
we will find that we need to choose one of the real {\it self-dual}
tensor fields to be non-vanishing at zeroeth order, to be able to
identify these linearized supergravity equations with the
constraints of the string model.

We will see that the two-form $b_{mn}$ is a linear combination of
{\it all} the 
oscillations corresponding to the five self-dual tensor
fields and the anti-self-dual tensor field,  including
the oscillation with non-vanishing background. 
In flat space, $b_{mn}$ corresponds to a state in the Neveu-Schwarz sector.	
In our curved space case,
the string model describes vertex operators for $\AdS_3$ background
with Ramond-Ramond flux. When matching the vertex operator component fields
with the supergravity oscillations, we find that not only the 
bispinor $V^{--}_{ab}$ (which is a Ramond-Ramond field in the
flat space case),  but also the tensor $b_{mn}$ 
include supergravity oscillations with non-vanishing self-dual background.
We use the Romans' variables \refs{\r}
to describe the $AdS$ supergravity equations, 
where the three-form field strengths are labelled by one anti-self-dual
$K_{mnp}$ and five self-dual $H^i_{mnp}$ fields. 
Our procedure will be to start in this section with the supergravity
equations, and rewrite them as equations for the field combinations that
occur in the string theory vertex operators. We then show that
these equations are equivalent to the string constraint equations
\ccone -\ccsix.
We list the field identifications at the end of this section. 

In the bosonic sector, the supergravity equations are
\eqn\sgone{\eqalign{&D^p D_p \phi^i = {\textstyle{2\over 3}}\,
H^i_{mnp}\, K^{mnp}\cr
&H^i_{mnp} = 
{\textstyle{1\over 6}} \, e_{mnp}^{\qquad qrs}\, H^i_{qrs}\,,
\qquad K_{mnp} = 
-{\textstyle{1\over 6}} \, e_{mnp}^{\qquad qrs}\, K_{qrs}\cr
&R_{mn} = - H^i_{mpq}\, H^{i pq}_{n} 
- K_{mpq}\, K^{pq}_{n} - D_m\phi^i \,D_n\phi^i\,.\cr}}  
where
$e_{mnpqrs}\equiv {1\over\sqrt{-\tilde g}}\,
\tilde g_{mm'}\ldots \tilde g_{ss'}\epsilon^{m'n'p'q'r's'}$  for a
metric $\tilde g$, and $\epsilon^{123456} = 1$, 
$1\le i\le 5$.  

The string vertex operator components are related to the 
supergravity fluctuations around the $\AdS_3\times \S^3$ metric
$\bar g_{mn}$ and the three-form field strength
$\bar H^i_{mpq} = \delta^{i1} \bar H^1_{mpq}$, 
where $\bar H^1_{mpq} = \half\,
(\s_m \s_p \s_q)_{ab}\delta^{ab}$, 
and $\bar g_{mn}$ is given in \Riemann.
As in \refs{\sez}, we can parametrize these
fluctuations as
\eqn\sgtwo{\eqalign{\tilde g_{mn} = \bar g_{mn} + h_{mn}\,,
\qquad H^i_{mpq} = \bar H^i_{mpq} + g^i_{mpq}\,,
\qquad K_{mpq} = g^6_{mpq} + \bar  H^j_{mpq} \phi^j\cr}}
where we label the scalar fluctuations by $\phi^i$ since they have
vanishing background. The anti-self-dual tensor $ K_{mpq}$
also has vanishing background, but is parametrized as above
so that the fluctuations $g^i_{mpq}$ and $g^6_{mpq}$ are exact 
three-forms \refs{\r}.

The linearized supergravity equations are given by 
\eqn\sglone{\eqalign{&D^p D_p \phi^i ={\textstyle{2\over 3}}\,
\bar H^i_{prs}\, g^{6prs} \cr}}
\eqn\sgltwo{\eqalign{&\half D^p D_p \, h_{rs}
- \bar R_{\tau r s \lambda} \, h^{\tau\lambda}
+ \half \bar R_r^\tau\, h_{\tau s}
+ \half \bar R_s^\tau\, h_{\tau r}
- \half D_s D^p h_{pr}
- \half D_s D^p h_{pr} + \half D_rD_s h^p_p\cr
&= 
\,- \bar H^{i\,\,\,pq}_r\, g^i_{spq}\,
- \bar H^{i\,\,\,pq}_s\, g^i_{rpq}\,
+ 2 h^{pt}\,\bar H^{i\,\,\,q}_{rp}\,\bar H^i_{stq}\cr}}
\eqn\sglthree{\eqalign{D^p H_{prs}
= &-2  \,\bar H^i_{prs}\, D^p\phi^i\cr
&+ B^i\,  [ - \bar H^{i\,\,pq}_r D_p\, h_{qs}
+ \bar H^{i\,\,pq}_s D_p\, h_{qr}
+ \bar H^{i\,\,\,q}_{rs} D^p h_{pq}
-{\textstyle{1\over 2}} 
\bar H^{i\,\,\,q}_{rs} D_q h^p_{\hskip5pt p}]\cr}}
where we have defined
$H_{prs} \equiv  \,g^6_{prs} + B^i \,g^i_{prs}$ 
as a combination of the supergravity exact forms $g^6\equiv db^6,
g^i\equiv d b^i$, since we will equate this with the string 
field strength $H=db$. For the moment we let $B^i$ be arbitrary constants.
In zeroeth order, the equations are
$\bar R_{rs} = - \bar H^i_{rpq}\,\bar H^{i\,\,pq}_s$.


\sglthree~ follows from the first order linearized duality equations
\eqn\sglfour{\eqalign
{{\textstyle{1\over 6}}\bar e_{mnp}^{\qquad qrs}\, g^6_{qrs}
=& - g^6_{mnp} - 2\phi^i \bar H^i_{mnp}\cr
{\textstyle{1\over 6}}\bar e_{mnp}^{\qquad qrs}\, g^i_{qrs}
=&  \,\,g^i_{mnp}
+ \bar H^i_{mpq}\, h_n^{\, q}
- \bar H^i_{npq}\, h_m^{\, q}
- \bar H^i_{mnq}\, h_p^{\, q}
+{\textstyle{1\over 2}}  \bar H^i_{mnp}  h_q^{\, q}\cr}}
with $\bar e_{mnpqrs}\equiv {1\over\sqrt{-\bar g}}\,
\bar g_{mm'}\ldots \bar g_{ss'}\epsilon^{m'n'p'q'r's'}$. 
We note that the fluctuations $g^6_{mnp}, g^i_{mnp}$
although exact, do not have definite duality due the presence of 
the metric.  

To equate the supergravity equations \sglone-\sglthree\ with
the string constraints of section 4, we remember that the
constraints hold for the string field combination
$G_{mn}\equiv g_{mn} + b_{mn} + \bar g_{mn}\phi$. 
To put \sglthree~ and \sgltwo~ in a similar form, and using
$H_{prs} \equiv \partial_p b_{rs} +  \partial_r b_{sp} +
\partial_s b_{pr}$,
we rewrite them as
\eqn\sglfive{\eqalign{{\textstyle{1\over 2}} D^p D_p b_{rs} =&
-   \bar H^1_{prs}\, D^p\phi^1
-\bar H^{1\,\,pq}_r \,\,[{\textstyle{1\over 2}}B^1\,D_p\, h_{qs}]
+\bar H^{1\,\,pq}_s \,\,[{\textstyle{1\over 2}}B^1\, D_p\, h_{qr}]\cr
&+ {\textstyle{1\over 2}} B^1 \,\,[\bar H^{1\,\,\,q}_{rs} D^p h_{pq}
-{\textstyle{1\over 2}} \bar H^{1\,\,\,q}_{rs} D_q h^p_{\hskip5pt p}]
\hskip10pt +\bar R_{\tau r s \lambda} \, b^{\tau\lambda}
-{\textstyle{1\over 2}} \bar R_r^{\,\,\tau}\, b_{\tau s}
-{\textstyle{1\over 2}} \bar R_s^{\,\,\tau}\, b_{r\tau}\cr
& +{\textstyle{1\over 2}}  D_r D^p b_{ps}
+{\textstyle{1\over 2}} D_s D^p b_{rp}\cr}}
\eqn\sglsix{\eqalign{{\textstyle{1\over 2}} D^p D_p \, h_{rs}=& 
-\bar H^{1\,\,\,pq}_r\,
\,[{2\over B^1} \,D_p b_{qs}] \,
+ \bar H^{1\,\,\,pq}_s\,
\,[{2\over B^1} \,D_p b_{rq}] \cr
&-\bar H^{1\,\,\,pq}_r\,
{1\over B^1} [\,- g^6_{spq} - B^I g^I_{spq}\,]
- \bar H^{1\,\,\,pq}_s\,
{1\over B^1} [\,- g^6_{rpq} - B^I g^I_{rpq}\,] \cr
&+ 2 h^{pt}\,\bar H^{1\,\,\,q}_{rp}\,\bar H^1_{stq}
+ \bar R_{\tau r s \lambda} \, h^{\tau\lambda}
\,-{\textstyle{1\over 2}} \bar R_r^{\,\,\tau}\, h_{\tau s}
-{\textstyle{1\over 2}} \bar R_s^{\,\,\tau}\, h_{r\tau }\cr
&+{\textstyle{1\over 2}}D_r D^p h_{ps}
+{\textstyle{1\over 2}}D_s D^p h_{rp}
-{\textstyle{1\over 2}}D_s D_r h^p_p\cr
&-D_r \bar H^{1\,\,\,pq}_s\,  
\,{1\over B^1} b_{pq} \,
-D_s \bar H^{1\,\,\,pq}_r\,
\,{1\over B^1} b_{pq}\cr}}
where $2\le I\le 5$.

The string graviton field $g_{rs}$ is chosen to be traceless and  
can be written in terms of the supergravity fields as
\eqn\sglsev{g_{rs} =  \,h_{rs} -{\textstyle{1\over 6}}\bar g_{rs}\,
h^\lambda_{\,\,\lambda}\,.}
Using \sglsev~ and the gauge conditions \gc~, we find \sglfive,\sglsix~ 
to be
\eqn\sgleight{\eqalign{{\textstyle{1\over 2}} D^p D_p b_{rs}
=&-\bar H^{1\,\,pq}_r \,
D_p\,[\,g_{qs} + \bar g_{qs}\phi\,]
+\bar H^{1\,\,pq}_s \,
\, D_p\,[\,g_{qr} + \bar g_{qr}\phi\,]\cr
&-\bar R_{\tau r s \lambda} \, b^{\tau\lambda}\,
-{\textstyle{1\over 2}} \bar R_r^{\,\,\tau}\, b_{\tau s}
-{\textstyle{1\over 2}} \bar R_s^{\,\,\tau}\, b_{r\tau}\cr
&+  \bar H^1_{prs}\, D^p\, [\,- \phi^1 \, - \, 3 \,\phi\,]\,,
\cr}}
\eqn\sglnine{\eqalign{{\textstyle{1\over 2}} D^p D_p \, (\, g_{rs} + \bar g_{rs} \phi \,)\,=
&-\bar H^{1\,\,\,pq}_r\,
D_p b_{qs} \,
+ \bar H^{1\,\,\,pq}_s\,
D_p b_{rq} \cr
&+ 2 (\, g^{pt} + \bar g^{pt}\phi\,)\,\bar H^{1\,\,\,q}_{rp}\,\bar H^1_{stq}\cr
&+ \bar R_{\tau r s \lambda} \,
(\, g^{\tau\lambda} + \bar g^{\tau\lambda}\phi \,)\,
-{\textstyle{1\over 2}} \bar R_r^{\,\,\tau}\,
(\, g_{\tau s} + \bar g_{\tau s} \phi \,)
-{\textstyle{1\over 2}} \bar R_s^{\,\,\tau}\,
(\, g_{r\tau } + \bar g_{r\tau }\phi \,)\cr
&-D_r D_s \phi\,
-{\textstyle{1\over 3}} D_r D_s h^\lambda_{\,\, \lambda}\cr
&+ \bar R_{rs}\, \,[\, -{\textstyle{1\over 3}} \,h^\lambda_{\,\, \lambda}
+ \phi^1\, +2\phi]\cr 
&-\bar g_{rs} \,D^p D_p\,
[\,{\textstyle{1\over 12}}\,h^\lambda_{\,\,\lambda}\,
- {\textstyle{1\over 4}}\,\phi^1 \,- {\textstyle{1\over 2}}\phi]\cr
&+\bar H^{1\,\,\,pq}_r\,
{\textstyle{1\over 2}} [ B^I g^I_{spq}\,]
+ \bar H^{1\,\,\,pq}_s\,
{\textstyle{1\over 2}} [ B^I g^I_{rpq}\,] 
\cr}}
where we have used \sglone~ and let 
$B^1 = 2$. Note that
$- {\textstyle{1\over 2}} \,\,\bar H^{1\,\,\,q}_{rs}
(\sigma_m\sigma^q\sigma_n)_{ab} \delta^{ab} \,\, b_{mn}
= -2\bar R_{\tau r s \lambda} \, b^{\tau\lambda}$.  
As discussed below, we make the field identifications:
\eqn\fione{\eqalign{&{\textstyle{1\over 4}} F^{++gh}_{\rm asy}
\,\sigma_r^{ab}\sigma_s^{ef}\, \delta_{ah}\delta_{be}\delta_{gf}
\equiv\bar H^1_{prs}\, D^p\, [\,- \phi^1 \, - 3\,\phi\,]\cr}}
\eqn\fitwo{\eqalign{
{\textstyle{1\over 4}} F^{++gh}_{\rm sym} \,\sigma_{rga}\sigma_{shb}\,
\delta^{ab}
\equiv &\,\,\bar H^{1\,\,\,pq}_r\,
{\textstyle{1\over 2}} [ B^I g^I_{spq}\,]
+ \bar H^{1\,\,\,pq}_s\,
{\textstyle{1\over 2}} [ B^I g^I_{rpq}\,]\cr
&-D_r D_s \phi\,
-{\textstyle{1\over 3}} D_r D_s h^\lambda_{\,\, \lambda}\cr
&+ \bar R_{rs}\, \,[\, -{\textstyle{1\over 3}} \,h^\lambda_{\,\, \lambda}
+ \,\phi^1\, +2\phi]\cr
&-\bar g_{rs} \,D^p D_p\,
[\,{\textstyle{1\over 12}}\,h^\lambda_{\,\,\lambda}\,
- {\textstyle{1\over 4}}\,\phi^1 \,- {\textstyle{1\over 2}}\phi]\cr}}
These identifications are appropriate for the following
reasons. 
{}From the trace of \fitwo~, 
$0 = D^p D_p \,[\,2\phi -{\textstyle{5\over 6}}
\,h^\lambda_{\,\,\lambda}\, +{\textstyle{3\over 2}} \phi^1\,]$,
and the trace of \sglsix~ 
\eqn\trone{ D^p D_p [\, {\textstyle{5\over 6}}\, h^\lambda_{\,\,\lambda}
+ \phi \, - {\textstyle{1\over 2}} \phi^1] = 0 \,,}
we find
\eqn\fithree{
D^p D_p\, [\,- \phi^1 \, - 3\,\phi\,]
= 0\,.}
\fithree, derived here from the supergravity equations, is also
the {\it string} equation \ssthree~ for $F^{++ab}_{\rm asy}$ derived in section 4 
{\it when } $F^{++ab}_{\rm asy}$ {\it is defined by} \fione. 

To make contact with the string equation \ssfour~ for $F^{++ab}_{\rm sym}$,
we multiply \fitwo~ by $\sigma^{s\,\,HB}\, D_r$, 
so that
\eqn\fifour{\eqalign{&
{\textstyle{1\over 4}}\,[\, \delta^{Ba} \,\sigma^r_{ga}\,D_r\,
F^{++gH}_{\rm sym} \, - \,
\delta^{Ha} \,\sigma^r_{ga}\,D_r\, F^{++gB}_{\rm sym} \, ]\cr
=\,\,&-{\textstyle{1\over 2}}\,\epsilon^{HB}_{\qquad cd}\,\,\sigma^{r\,cd}\, D_r
\,[\,  \phi^1 + 3\phi]\cr
=\,\,&- {\textstyle{1\over 4}}\,\, \epsilon^{BH}_{\qquad cd}\,\,
F^{++cd}_{\rm asy}\cr}}
where we have used the equations of motion $D^p g_{pqs}^I = 0$, $2\le I\le 5$.
{}From \sglone, the equations for the scalars are
\eqn\fifive{\eqalign{&D^p D_p \phi^I = 0\,,\qquad {2\le{\rm I}\le 5}\cr
&D^p D_p \,\phi^1
= {\textstyle{2\over 3}} \bar H^{1prs}\, H_{prs} \,
+ 2 \bar R_{rp} \, g^{rp}\cr}}
{}From \trone,\fifive~ we make the field identification between the
string dilaton field $\phi$ and the supergravity scalars
$h^\lambda_{\,\,\lambda},\phi^i$:
\eqn\fisix{\phi \equiv {\textstyle{1\over 2}} \,\phi^1       
- {\textstyle{5\over 6}} h^\lambda_{\,\,\lambda}
\, + C^I\phi^I\,.}
Note that for $\AdS_3\times \S^3$ we have
\eqn\fisev{ 2 (\, g^{pt} + \bar g^{pt}\phi\,)\,\bar H^{1\,\,\,q}_{rp}\,\bar H^1_{stq}
= -2  \bar R_{\tau r s \lambda} \,
(\, g^{\tau\lambda} + \bar g^{\tau\lambda}\phi \,)\,.}
With these field identifications, we can combine \sgleight,\sglnine~
to find from the supergravity that
\eqn\fieight{\eqalign{{\textstyle{1\over 2}} D^p D_p G_{rs} =&
-\bar H^{1\,\,pq}_r \,\, D_p\, G_{qs} \,
+\bar H^{1\,\,pq}_s \,\, D_p\, \,G_{rq} \,\cr 
&-\bar R_{\tau r s \lambda} \, G^{\tau\lambda}
-{\textstyle{1\over 2}} \bar R_r^{\,\,\tau}\, G_{\tau s}
-{\textstyle{1\over 2}} \bar R_s^{\,\,\tau}\, G_{r\tau}\cr
&+ {\textstyle{1\over 4}} F^{++gh}_{\rm asy}
\,\sigma_r^{ab}\sigma_s^{ef}\, \delta_{ah}\delta_{be}\delta_{gf}\cr
&+{\textstyle{1\over 4}} F^{++gh}_{\rm sym} \,\sigma_{rga}\sigma_{shb}\,
\delta^{ab}\cr}}
which is the zero Laplacian string conditions \ssone,\sstwo.

For the string constraints on $V^{--}_{ab}$, 
we have from the trace of \ssfive
\eqn\finine{\eqalign{8\,\bar R_{mn}\, g^{mn}&=
{\textstyle{1\over 2}} \, D^p D_p\,\delta^{cd}\, V_{cd}^{--}
\, +\, D^m\sigma_{mab}\,\delta^{ac}\delta^{bd}\, V_{cd}^{--}\,,\cr}}
and acting on \ssfive~ with $(\sigma^{rcd} D_r - \delta^{cd})$ 
and using the supergravity equations \fifive~ we find
\eqn\fiten{\eqalign{-8 \phi^1 \,+\, 8 C^I_{--}\phi^I 
&= \sigma^{ncd}\, D_n\, V_{cd}^{--} \,
-2\,\delta^{cd}\, V_{cd}^{--}\,.\cr}}
{}From \ssfive~ and the second order equation for $g_{qrs}^i$ which follows
from \sglfour~, one can show that 
$(\,\sigma^{mac}\,\sigma^{nbd}\, D_m\,D_n \, V_{cd}^{--}\,)_{\rm sym}$
involves a combination of the supergravity fields
$ (\sigma_p\sigma_m\sigma_n)^{ab} \, g^1_{pmn}$ and
$ (\sigma_p\sigma_m\sigma_n)^{ab} \, B_{--}^I g_{pmn}^I.$

To summarize the  vertex operator components 
in terms of the supergravity fields
$g^i_{prs}, g^6_{prs}, h_{rs}, \phi^i$ , $1\le i\le 5$, (and $2\le I\le 5$ below):
\eqn\one{H_{prs}\equiv  \,g^6_{prs} + 2 \,g^1_{prs}\,+ B^I \,g^I_{prs}}
\eqn\two{g_{rs} \equiv \,h_{rs}
-{\textstyle{1\over 6}}\bar g_{rs}\, h^\lambda_{\,\,\lambda}}
\eqn\three{\phi = -{\textstyle{1\over 3}}\, h^\lambda_{\,\,\lambda}} 
\eqn\five{\eqalign{F^{++ab}_{\rm sym} =  &\,\, {\textstyle{2\over 3}}
(\sigma_p\sigma_r\sigma_s)^{ab}
\, B^I \,\,g^I_{prs} +  \delta^{ab}\,\phi^{++}\cr
F^{++ab}_{\rm asy} = &\,\,\sigma^{p\,ab} \, D_p\,\phi^{++}\cr
\phi^{++} &=  4 C^I\,\,\phi^I\cr}} 
which follows from choosing the graviton trace
$h^\lambda_{\hskip3pt\lambda}$ to satisfy $\phi^1 -
h^\lambda_{\,\,\lambda} \,\equiv\, - 2\, C^I\phi^I$.
This plays the role of the harmonic coordinate condition
which occurs in flat space string amplitudes. 

The combinations $C^I\phi^I$ and $B^I g^I_{prs}$ reflect the
$SO(4)_{\rm R}$ symmetry of the $D=6, N=(2,0)$ theory on $\AdS_3\times \S^3$.
We relabel $C^I = C^I_{++}$, $B^I = B^I_{++}$.
To define the remaining string components in terms of supergravity fields,
we consider linearly independent quantities $C_\ell^I\phi^I$,
$B_\ell^I  g^I_{prs}$, $\ell = ++,+-, -+, --$. 
\eqn\five{\eqalign{F^{+-ab}_{\rm sym} =  &\,\, {\textstyle{2\over 3}}
(\sigma_p\sigma_r\sigma_s)^{ab}
\, B_{+-}^I \,\,g^I_{prs} +  \delta^{ab}\,\phi^{+-}\cr
F^{+-ab}_{\rm asy} = &\,\,\sigma^{p\,ab} \, D_p\,\phi^{+-}\cr
\phi^{+-} = & 4 C_{+-}^I\,\,\phi^I\cr
F^{-+ab}_{\rm sym} =  &\,\, {\textstyle{2\over 3}}
(\sigma_p\sigma_r\sigma_s)^{ab}
\, B_{-+}^I \,\,g^I_{prs} +  \delta^{ab}\,\phi^{-+}\cr
F^{-+ab}_{\rm asy} = &\,\,\sigma^{p\,ab} \, D_p\,\phi^{-+}\cr
\phi^{-+} =& 4 C_{-+}^I\,\,\phi^I\cr}}
$V^{--}_{ab}$ is given in terms of the fourth tensor/scalar pair
$C_{--}^I\,\,\phi^I$, $B_{--}^I g_{mnp}^I$ through
\eqn\six{\eqalign{&
D^p D_p\, V_{cd}^{--}
\, - \delta^{gh}\sigma^p_{ch}\, D_p \, V_{gd}^{--}
\, + \delta^{gh}\sigma^p_{dh}\, D_p \, V_{cg}^{--}
\, +{\textstyle{1\over 2}} \epsilon_{cd}^{\,\,\,\,gh}\, V_{gh}^{--}
=\,-8\,\sigma^m_{ce}\,\sigma^n_{df}\,\delta^{ef}\, G_{mn}\,.\cr}} 

The fermion constraints in \ccfer~
imply the linearized AdS supergravity equations for the gravitinos
and spinors, due to the above correspondence for the bosons and the
supersymmetry of the two theories. 

\vskip50pt
{\bf Acknowledgements:}
We would like to thank Nathan Berkovits for discussions.  LD thanks the 
Institute for Advanced Study for its hospitality, 
and was partially supported by the  U.S. Department of Energy,
Grant No. DE-FG 05-85ER40219/Task A. Research of EW was supported in part by
NSF Grant PHY-9513835 and the Caltech Discovery Fund.

\vfill\eject
\appendix {A} {\hskip10pt Sigma Matrices}

The sigma matrices $\s^{m ab}$ \refs{\bvw,\gsw} satisfy the algebra
\eqn\sig{\eqalign{\s^{m ab} \s^n_{ac} + \s^{n ab} \s^m_{ac} 
&= \eta^{mn} \delta^b_c\cr}} 
in flat space, where $\eta^{mn}$ is the six-dimensional Minkowski metric,
and $1\le a\le 4$.
Sigma matrices with lowered indices are defined by
$\s^m_{ab} = \half\e_{abcd} \s^{m cd}$, although for other
quantities indices are raised and lowered with $\delta^{ab}$,
so we distinguish $\s^m_{ab}$ from $\delta_{ac}\,\delta_{bd}\,\s^{m cd}$.
The following identities are useful
\eqn\sigid{\eqalign 
{\s^{m ab} \s^n_{cd} \,\eta_{mn} = 
\delta^a_c \delta^b_d - \delta^a_d \delta^b_c\,,\qquad
\s^{m ab} \s^{n cd} \,\eta_{mn}  = \e^{abcd}\,.\cr}}
Also \sig,\sigid~ hold in curved space with $\eta_{mn}$ replaced
by the $\AdS_3\times \S^3$ metric $\bar g_{mn}$.
Then $\bar g^{mn} = \half \s^{m ab} \s^n_{ab}$ as in \Riemann.

The product of sigma matrices 
$<G_{mnp}> \equiv (\sigma_m\sigma_n\sigma_p)_{ab} \,\delta^{ab}$
is self-dual.
Similarly $(\sigma_m\sigma_n\sigma_p)^{ab} \,\delta_{ab}$ 
is anti-self-dual.
The sigma matrices $\s^{m ab}$ describe the coupling of two
{\bf 4}'s to a {\bf 6} of $SU(4)$. They are $SU(4)$ singlets,
whereas $\delta^{ab}$, though an $SO(4)$ singlet,
transforms in  the symmetric second rank tensor {\bf 10} of $SU(4)$.
{}From the group theory properties of the sigma matrices, we
see that  
$<G_{mnp}>$ is a {\bf 10} of $SU(4)$, so the product of the
two self-dual tensors, which is a singlet of $SU(4)$, must be zero 
$<G^{mnp}> \,<G_{mnp}> = 0$, since $ {\bf 10} \times  {\bf 10} =
{\bf 20'} + {\bf 35}  + {\bf 45} $ does not contain a singlet.  
This is consistent with the general identities  \r\ that
\eqn\sdid{\eqalign{X_{[m}^{\,np} Y_{r ]\,np} = 0 =
X^{mnp} \, Y_{mnp}\cr}}
when the tensors $X,Y$ are both self-dual (or both anti-self-dual);
and
\eqn\asdid{\eqalign{X_{(m}^{\,np} Z_{r )\,np} = 
{\textstyle{1\over 6}} g_{mr} \, X^{nps} \, Z_{nps}\,\cr}} 
for $X,Z$ of opposite duality.

\vfill\eject
\appendix {B} {\hskip10pt Covariant Derivatives on a Group Manifold}

On a group manifold, there is always a right-invariant vielbein
and a left-invariant vielbein; for the $SO(4)$ case, we write them as
\eqn\vb{\eqalign{e^{m ab}= - \s^{m ab}\cr}}
and
\eqn\lvb{e^{m \bar a\bar b} = \s^{m \bar a\bar b}}
respectively.
The vielbein obeys 
\eqn\vbtwo{\eqalign{e^{m ab}\, e_{m}^{cd} &= \epsilon^{abcd}\,,
\qquad e^{m ab}\, e^{n cd}\, \e_{abcd} = 4 \,\bar g^{mn}\,.\cr}}  

Using the right-invariant vielbein, the covariant derivative
on the group manifold can be conveniently written
\eqn\bcovd{\eqalign{D_m V_{gh} & = \p_m  V_{gh}
-{\textstyle{1\over 4}}\, f_{gh\quad ef}^{\quad cd}\,e_m^{ef}\,  V_{cd}\cr}}
for $V_{gh}$ antisymmetric in g,h.
Here by $f$ we mean the structure constants $f^A_{BC}$, with $A,B, $ and
$C$ being Lie algebra indices, but
for the group $SO(4)$, the Lie algebra is the second rank antisymmetric
tensor representation, and we write a Lie algebra index as a pair $ab$
with antisymmetry understood.
It is convenient to refer the covariant derivative to the right-invariant
vielbein, defining ${\cal T}_L^{ab}=e^{mab}D_m$.  Similarly,
we define ${\cal T}_R^{\bar a\bar b}=e^{m\bar a\bar b}D_m$, using
the left-invariant vielbein.

It is also convenient to let $t_L$ be a covariant derivative
that is defined by using the right-invariant vielbein and setting
the spin connection to zero; likewise, $t_R$ is a covariant
derivative defined using the left-invariant vielbein and setting
the spin connection to zero.  Thus, $t_L$ and $t_R$ are the covariant
derivatives defined using the right- and left-invariant framings 
of the group manifold.

{}From the above formulas, one can deduce the relation between $t_L$
and ${\cal T}_L$.  On a Lie algebra valued field $V_{gh}$ ($=-V_{hg}$)
we have
\eqn\lfttwo{\eqalign{t_L^{ab} \,  V_{gh} &= {\cal T}_L^{ab} \,  V_{gh} 
+ \half f_{gh}^{\,\,\,\,{cd ab}}  V_{cd}\,.\cr}} 
{}From this, we can deduce that on a spinor field the relation is 
\eqn\oneinvd{\eqalign{t_L^{ab} V_g &=
{\cal T}_L^{ab} \, V_g + \half \delta^a_g\, \delta^{bc} V_c
- \half \delta^b_g\, \delta^{ac} V_c\,\cr}} 
and on a field with two pairs of anti-symmetric indices it is
\eqn\oneinvd{\eqalign{t_L^{ab} V_{gh\,\, jk} &= 
{\cal T}_L^{ab} \, V_{gh\,\, jk} +
\half f_{gh}^{\,\,\,\,{cd ab}}  V_{cd\,\, jk}\,
+\half f_{jk}^{\,\,\,\,{cd ab}}  V_{gh\,\,cd}\,.\cr}}

The above formulas have obvious counterparts for $t_R$ and ${\cal T}_R$.

It is convenient, as in the case of the sigma matrices, 
to raise and lower the index pairs on the structure
constants with $\epsilon_{abcd}$, so $\half\epsilon_{abef}
f_{gh}^{\,\,cdab}= f_{gh\quad ef}^{\quad cd}$. Otherwise indices
are raised and lowered with $\delta^{ab}$.

\listrefs

\bye